\documentclass[twocolumn,showpacs,fleqn,nobibnotes]{revtex4}

\usepackage{amsmath}
\usepackage{graphicx}
\usepackage{float}
\usepackage{subfigure}

\def\lsim{\raise0.3ex\hbox{$<$\kern-0.75em\raise-1.1ex\hbox{$\sim$}}}
\def\gsim{\raise0.3ex\hbox{$>$\kern-0.75em\raise-1.1ex\hbox{$\sim$}}}

\renewcommand{\vec}[1]{\boldsymbol{#1}}
\newcommand{\dif}{\mathrm{d}}

\begin{document}

\title{Investigating the exclusive protoproduction of dileptons at high energies}
\author{Magno V.T. Machado}
\affiliation{Centro de Ci\^encias Exatas e Tecnol\'ogicas, Universidade Federal do Pampa \\
Campus de Bag\'e, Rua Carlos Barbosa. CEP 96400-970. Bag\'e, RS, Brazil}

\begin{abstract}
Using the high energy color dipole approach, we study the exclusive photoproduction of lepton pairs, $\gamma N \rightarrow \gamma^*(\rightarrow \ell^+\ell^-)\, N$ (with $N=p,\,A$). We use simple models for the elementary dipole-hadron scattering amplitude that captures main features of the dependence on atomic number $A$, on energy and on momentum transfer $t$. This investigation is complementary to conventional partonic description of timelike Compton scattering, which considers quark handbag diagrams at leading order in $\alpha_s$ and simple models of the relevant generalized  parton distributions (GPDs). These calculations are input in electromagnetic interactions in $pp$ and $AA$ collisions to measured at the LHC.

\end{abstract}

\maketitle

\section{Introduction}

Recently, the physics of large impact parameter interactions at the LHC and Tevatron has raised great interest (see Refs. \cite{YRUPC} for a review). These electromagnetic interactions in $pp$ and $AA$ collisions extend the physics program of photon induced processes beyond the energies currently reached at DESY-HERA. At very high collision energies, the electromagnetic field surrounding a nucleus contains photons energetic enough to
produce new particles in ultra-peripheral collisions. This can happen in a
purely electromagnetic process through a two-photon interactions or in an
interaction between a photon from one of the nuclei and the other ``target'' nucleus. The high photon energies and fluxes lead to large cross sections for several photon-induced reactions.  The coherent contribution from the $Z$ protons in the nucleus, furthermore, enhances the number of equivalent photons by a factor $Z^2$.

These ultraperipheral collisions (UPCs) are a good place to constraint the photonuclear cross sections as the dominant processes in UPCs are photon-nucleon (nucleus) interactions. The current LHC detector configurations can explore small-$x$ hard phenomena with nuclei and nucleons at photon-nucleon center-of-mass energies above 1 TeV, extending the $x$ range of HERA by a factor of ten. In particular, it will be possible to probe diffractive and inclusive parton densities in nuclei using several processes. For instance, the interaction of small dipoles with protons and nuclei can be investigated in elastic and quasi-elastic $J/\psi$ and $\Upsilon$ production as well as in high $t$ $\rho^0$ production accompanied by a rapidity gap. Several of these phenomena provide clean signatures of the onset of the new high gluon density QCD regime \cite{Strik}. The LHC is in the kinematic range where nonlinear effects are several times larger than at HERA.

Electromagnetic interactions can also be studied with beams of
protons or anti-protons, but there is then no $Z^2$-enhancement in the
photon flux in contrast to $AA$ collisions. Several analysis are currently being done at Tevatron focusing on such processes. For instance, CDF Collaboration is analyzing the exclusive production of muon pairs,
$p \overline{p} \rightarrow p \overline{p} + \mu^+ \mu^-$, at lower invariant
masses \cite{Confpress}. The two main contributions to these events are, as with
heavy-ion beams, $\gamma \gamma \rightarrow \mu^+ + \mu^-$ and
$\gamma + I\!P\rightarrow J/\Psi (\mathrm{or} \,\Psi') $, followed by the meson decay
into a dilepton pair.  The continuum $\gamma + \gamma \rightarrow \mu^+ \mu^-$
has been calculated \cite{Nystranddilep} under the same conditions as the vector mesons photoproduction and it was found a cross section, $p\overline{p} \rightarrow p\overline{p} + \mu^+\mu^-= 2.4$ nb. In that calculation, it was used the cut $M_{\mathrm{inv}} >$~1.5 GeV for the dilepton invariant mass.

In this work, we use the high energy color dipole approach \cite{dipole} to study the exclusive photoproduction of lepton pairs, $\gamma N \rightarrow \gamma^*(\rightarrow \ell^+\ell^-)\, N$ (with $N=p,\,A$). The main reason is that such a process should be a background for the main reaction $\gamma \gamma \rightarrow \ell^+\ell^-$.  We use simple models for the elementary dipole-hadron scattering amplitude that captures main features of the dependence on atomic number $A$, on energy and on momentum transfer $t$. This investigation is somewhat complementary to conventional partonic description of timelike Compton scattering (TCS) \cite{Diehl}, $\gamma p \rightarrow \gamma^*+p$, which considers the relevant generalized  parton distributions (GPDs).  It should be noticed that the TCS process has so far only been studied at LO in the collinear factorization framework in terms of the quark GPDs and sub-processes initiated by gluons have not been considered \cite{Diehl}. The color dipole approach provides a very good description of the data on $\gamma p$ inclusive production, $\gamma \gamma$ processes, diffractive deep inelastic and vector meson production (For a recent review, see \cite{fss}). In particular, the electromagnetic deeply virtual Compton scattering (DVCS) cross section is nicely reproduced in several implementations of the dipole cross section at low $x$ \cite{dvcs1,FM,KMW,MPS,Watt,MW}. The TCS process at small $t$ and large timelike virtuality of the outgoing photon shares many features of DVCS.

This paper is organized as follows.  In next section, we make a short summary on the color dipole approach applied to the exclusive protoproduction of dileptons at high energies. In Sec.  3, we present the main results for such a process using proton and nucleus targets. For sake of simplicity in numerical calculations, we consider the space-like kinematics in this preliminary estimation of relevant cross sections. Finally, in last section we summarize the results.

\section{TCS process within the color dipole approach}

Let us summarize the relevant formulas in the color dipole picture for the  TCS process. In this formalism \cite{dipole}, the scattering process $\gamma p\rightarrow \gamma^*p$ is assumed to proceed in three stages: first the incoming real photon fluctuates into a quark--antiquark pair, then the $q\bar{q}$ pair scatters elastically on the proton, and finally the $q\bar{q}$ pair recombines to form a virtual photon (which subsequently decays into lepton pairs). The amplitude for production of the exclusive final state such as a virtual photon in TCS, is given by \cite{MPS,KMW,MW}
\begin{eqnarray}
 \mathcal{A}^{\gamma p\rightarrow \gamma^* p}(x,Q,\Delta) & = & \sum_f \sum_{h,\bar h} \int\!\dif^2\vec{r}\,\int_0^1\!\dif{z}\,\Psi^*_{h\bar h}(r,z,Q)\nonumber \\
& \times & \mathcal{A}_{q\bar q}(x,r,\Delta)\,\Psi_{h\bar h}(r,z,0)\,,
  \label{eq:ampvecm}
\end{eqnarray}
where $\Psi_{h\bar h}(r,z,Q)$ denotes the amplitude for a photon to fluctuate into a quark--antiquark dipole with helicities $h$ and $\bar h$ and flavour $f$. The quantity $\mathcal{A}_{q\bar q}(x,r,\Delta)$ is the elementary amplitude for the scattering of a dipole of size $\vec{r}$ on the proton, $\vec{\Delta}$ denotes the transverse momentum lost by the outgoing proton (with $t=-\Delta^2$), $x$ is the Bjorken variable and $Q^2$ is the photon virtuality. The elementary elastic amplitude $\mathcal{A}_{q\bar q}$ can be related to the $S$-matrix element $S(x,r,b)$ for the scattering of a dipole of size $\vec{r}$ at impact parameter $\vec{b}$ \cite{MPS,KMW}:
\begin{eqnarray}
  \mathcal{A}_{q\bar q}(x,r,\Delta) = \mathrm{i}\,\int \dif^2\vec{b}\;\mathrm{e}^{-\mathrm{i}\vec{b}\cdot\vec{\Delta}}\,2\left[1-S(x,r,b)\right].
  \label{eq:smatrix}
\end{eqnarray}

As one has a real photon at the initial state, only the transversely polarized overlap function contributes to the cross section.  Summed over the quark helicities, for a given quark flavour $f$ it is given by \cite{MW},
\begin{eqnarray}
  (\Psi_{\gamma^*}^*\Psi_{\gamma})_{T}^f & = & \frac{N_c\,\alpha_{\mathrm{em}}e_f^2}{2\pi^2}\left\{\left[z^2+\bar{z}^2\right]\varepsilon_1 K_1(\varepsilon_1 r) \varepsilon_2 K_1(\varepsilon_2 r) \right.\nonumber \\
& + &  \left. m_f^2 K_0(\varepsilon_1 r) K_0(\varepsilon_2 r)\right\},
  \label{eq:overlap_dvcs}
\end{eqnarray}
where we have defined the quantities $\varepsilon_{1,2}^2 = z\bar{z}\,Q_{1,2}^2+m_f^2$ and $\bar{z}=(1-z)$. Accordingly, the photon virtualities are $Q_1^2=0$ (incoming real photon) and $Q_2^2=-Q^2$ (outgoing virtual photon).  For sake of simplicity, in numerical calculations we use space-like kinematics. We expect not a large deviation from the correct kinematics. However, the approximation we have considered to estimate the cross section should be taken with due care. We quote Ref. \cite{MW} for a detailed derivation of wavefunctions for time-like photons.  The elastic diffractive cross section is then given by (disregarding the real part of amplitude),
\begin{eqnarray}
  \frac{\dif\sigma^{\gamma p\rightarrow \gamma^* p}}{\dif t}
  & = & \frac{1}{16\pi}\left\lvert\mathcal{A}^{\gamma p\rightarrow \gamma^* p}(x,Q,\Delta)\right\rvert^2
  \label{eq:xvecm1}
\end{eqnarray}

It should be noticed that some corrections to this exclusive process are needed. For TCS one should use the off-diagonal gluon distribution, since the exchanged gluons carry different fractions $x$ and $x^\prime$ of the proton's (light-cone) momentum. The skewed effect can be accounted for, in the limit that $x^\prime \ll x \ll 1$, by multiplying the elastic differential cross section by a factor $R_g^2$, given by \cite{Shuvaev:1999ce}
\begin{eqnarray}
\label{eq:Rg}
  R_g(\lambda) = \frac{2^{2\lambda+3}}{\sqrt{\pi}}\frac{\Gamma(\lambda+5/2)}{\Gamma(\lambda+4)},\! \quad\text{with}\!\quad \lambda \equiv \frac{\partial\ln\left[\mathcal{A}(x,Q^2,|t|)\right]}{\partial\ln(1/x)}.\nonumber
\end{eqnarray}

In our numerical calculations we take into account saturation models which successfully describe exclusive processes at high energies (vector meson production, diffractive DIS and DVCS). As a baseline model we consider the non-forward saturation model of Ref. \cite{MPS} (hereafter MPS model). It has the great advantage of giving directly the $t$ dependence of elastic differential cross section without the necessity of considerations about the impact parameter details of the process. Furthermore, the overall normalization of the total cross section is completely determined and no assumption about elastic slope is needed. Therefore, this simple saturation model captures main features of the dependence on energy,  virtual photon virtuality and momentum transfer $t$. An important result about the growth of the dipole amplitude towards the saturation regime is the geometric scaling regime \cite{gsincl,travwaves}. It first appeared in the context of the proton structure function, which involves the
dipole scattering amplitude at zero momentum transfer. At small values of $x,$
instead of being a function of both the variables $r$
and $x$, the dipole scattering amplitude is actually a function of the single
variable $r^2Q^2_{\mathrm{sat}}(x)$ up to inverse dipole sizes significantly larger than the
saturation scale $Q_{\mathrm{sat}}(x).$ More precisely, one can write
\begin{eqnarray}
\label{eq:geomsc0}
\mathcal{A}_{q\bar q}(x,r,\Delta=0)=2\pi R_p^2\: N\left(r^2Q_{\mathrm{sat}}^2(x)\right)\ ,
\end{eqnarray}
implying the geometric scaling of the total cross-section
at small $x$, i.e. $\sigma^{\gamma^*p\rightarrow X}_{\text{tot}}(x,Q^2)=
\sigma^{\gamma^*p\rightarrow X}_{\text{tot}}(\tau_p\!=\!Q^2/Q_{\mathrm{sat}}^2).$ This geometric scaling property can be extended to the case
of non zero momentum transfer \cite{MAPESO}, provided $r\Delta\ll 1$. Therefore, it has been obtained that
equation \eqref{eq:geomsc0} can be generalized to
\begin{eqnarray}
\label{sigdipt}
\mathcal{A}_{q\bar q}(x,r,\Delta)= 2\pi R_p^2\,e^{-B|t|}N \left(rQ_{\mathrm{sat}}(x,|t|),x\right),
\end{eqnarray}
with the asymptotic behaviors $Q_{\mathrm{sat}}^2(x,\Delta)\sim
\max(Q_0^2,\Delta^2)\,\exp[-\lambda \ln(x)]$. Specifically, the $t$ dependence of the saturation scale is parametrised as
\begin{eqnarray}
\label{qsatt}
Q_{\mathrm{sat}}^2\,(x,|t|)=Q_0^2(1+c|t|)\:\left(\frac{1}{x}\right)^{\lambda}\,, \end{eqnarray}
in order to interpolate smoothly between the small and intermediate transfer
regions. The form factor $F(\Delta)=\exp(-B|t|)$ catches the transfer dependence of the proton vertex, which is factorised from the
projectile vertices and  does not spoil the geometric scaling properties. Finally, the scaling function $N$ is obtained from the forward saturation model
\cite{Iancu:2003ge}, whose functional form is given by,
\begin{equation}
\label{eq:bcgc}
 N(x,\,r) =\begin{cases}
  \mathcal{N}_0\left(\frac{rQ_{\mathrm{sat}}}{2}\right)^{2\left(\gamma_s+\frac{1}{\kappa\lambda Y}\ln\frac{2}{rQ_{\mathrm{sat}}}\right)} & :\quad rQ_{\mathrm{sat}}\le 2\\
  1-\mathrm{e}^{-A\ln^2(BrQ_{\mathrm{sat}})} & :\quad rQ_{\mathrm{sat}}>2
  \end{cases},
\end{equation}
where $Y=\ln(1/x)$. The parameters for $N$ are taken from the original CGC model \cite{Iancu:2003ge}.

\section{Phenomenological results for TCS at high energies}

In this section the numerical results for the TCS cross section are presented using the color dipole picture and using the MPS saturation model as a representative phenomenological model. The sensitivity to a different choice for the dipole cross section is also analyzed. Here, we investigate the exclusive photoproduction of a heavy timelike photon which decays into a lepton pair, $\gamma p \rightarrow \ell^+\ell^-p$. Therefore, for the $\ell^+\ell^-$ invariant mass distribution from the virtual $\gamma^*$ decay we have (with $Q^2=M_{\ell^+\ell^-}^2$),
\begin{eqnarray}
 \frac{d\sigma }{dM_{\ell^+\ell^-}^2}\left(\gamma p\rightarrow \ell^+\ell^- p\right) = \frac{\alpha_{em}}{3\pi M_{\ell^+\ell^-}^2}\,\sigma\left(\gamma p \rightarrow \gamma^* p \right).
\end{eqnarray}

\begin{figure}[t]
\includegraphics[scale=0.47]{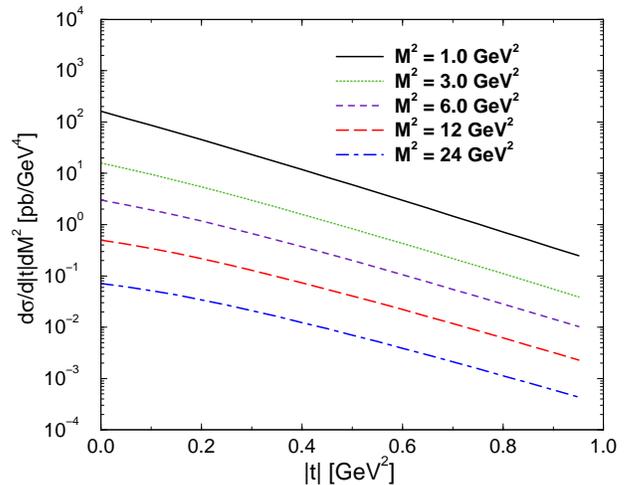}
\caption{(Color online) The differential cross section $d^2\sigma /d|t|dM^2$ as a function of $|t|$ at fixed values of dilepton invariant mass.}
\label{fig:1}
\end{figure}

In Fig. \ref{fig:1}  the differential cross section, $d^2\sigma/d|t|dM^2$, is presented as a function the momentum transfer (with $|t|<1$) for representative values of the dilepton invariant mass. We consider the typical DESY-HERA energy, $W_{\gamma p}=82$ GeV, where the  DVCS cross section is currently measured.  From a qualitative point of view, at small $|t|$ the differential cross section is consistent with an exponential behavior and a $M^2$-dependent slope. The general $|t|$ dependence is determined by the elementary dipole-nucleon scattering amplitude. For small size dipoles, one can qualitatively compute the scattering amplitude, which gives ${\cal A}\propto (Q_{\mathrm{sat}}^2/M^2)$. For the non-forward saturation model, $Q_{\mathrm{sat}}^2 \propto (1+c|t|)\exp(-B|t|)$. Thus, we checked that the cross section can be parameterized as $d\sigma/dtdM^2=a\,(1+c\,|t|)\exp(-b|t|)$, with the constants $a,\,c$ and $b$ depending logarithmically on $M^2$, $b=b(M^2)$. It should be noticed that the $|t|$-dependence is distinct for other implementations of the dipole cross section. However, for integrated cross sections, these deviations probably do not play an important role.

\begin{figure}[t]
\includegraphics[scale=0.47]{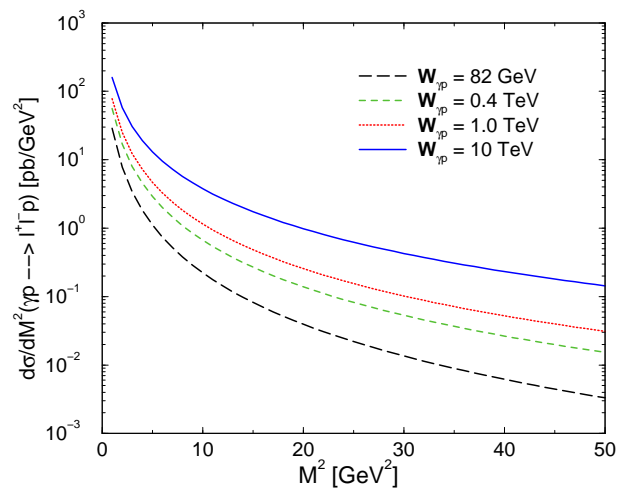}
\caption{(Color online) The differential cross section, $d\sigma/dM^2$, as a function of dilepton invariant mass for fixed values of energy (integrated over $|t|<1$ GeV$^2$).}
\label{fig:2}
\end{figure}

In the following we compute the integrated cross section, performing the integration over $|t|\leq 1$ GeV$^2$. In Fig. \ref{fig:2} we focus on the invariant mass dependence for fixed values of energy. In order to test the energy range which can be covered in electromagnetic interactions in $pp$ collisions, we start with the typical DESY-HERA value, $W=82$ GeV, and extrapolate it up to $W=10$ TeV.  Accordingly, the spectrum is dominated by small invariant masses of order $M_{\ell^+\ell^-}^2<20$ GeV$^2$, independent of energy. In the saturation models, the qualitative behavior can be obtained considering dominance of small size dipoles configuration.  The forward amplitude reads as ${\cal A}\propto (Q_{\mathrm{sat}}^2/M^2)$ times a logarithmic enhancement on $M^2$. Therefore, qualitatively the differential cross section behaves like $d\sigma/dM^2\propto (M^2)^{-\delta}[1+\log(M^2)]$, with $\delta \approx 3$.

\begin{figure}[t]
\includegraphics[scale=0.47]{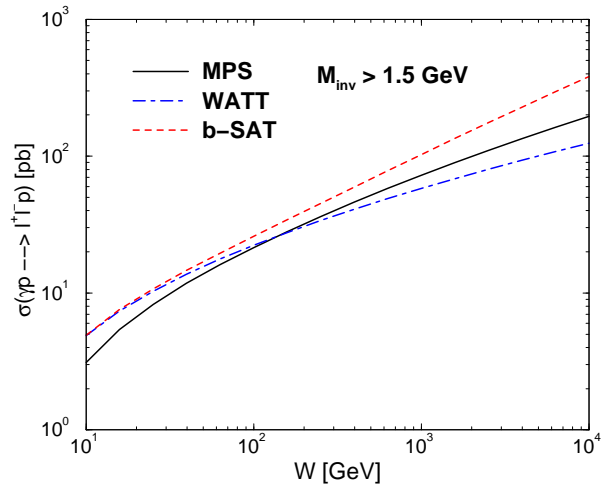}
\caption{(Color online)  The TCS integrated cross section ($M_{\ell^+\ell^-}\geq 1.5$ GeV and $|t|\leq 1$ GeV$^2$) as a function of photon-nucleus centre of mass energy (see text).}
\label{fig:3}
\end{figure}

In Fig. \ref{fig:3} we compute the total cross section, $\sigma(\gamma p \rightarrow \ell^+\ell^-\,p)$, integrated over dilepton invariant mass $M_{\ell^+\ell^-}\geq 1.5 $ GeV. In this plot, we compare the MPS model (solid line) with other implementations of the elementary dipole-nucleon scattering amplitude. For energies $W>>100$ GeV, a power fit can be performed in the form $\sigma = A\,(W/W_0)^\alpha$, with $W_0=1$ GeV. Considering the MPS model, one obtains the values $A= 3$ pb and $\alpha = 0.46$. In order to study the sensitivity to the model dependence, we compare the MPS saturation model to two different distinct saturation models. The first one is the recent implementation of the impact parameter Color Glass Condensate model \cite{Watt} (hereafter WATT, dot-dashed line). The second one is the impact parameter saturation model \cite{KMW}(hereafter b-SAT, long dashed line). In these implementations, the elementary dipole-nucleon scattering amplitude is written in the impact parameter space as referred in Eq. (\ref{eq:smatrix}). In particular, in the b-SAT model the $S$-matrix element is given by \cite{KMW},
\begin{eqnarray}
S(x,r,b)=\exp\left[-\frac{\pi^2}{2N_c}r^2\alpha_S(\mu^2)\,xg(x,\mu^2)\,T(b)\right]\nonumber  \\
\label{bsatelem}
\end{eqnarray}

Here, the scale $\mu^2$ is related to the dipole size $r$ by $\mu^2=4/r^2+\mu_0^2$.  The gluon density, $xg(x,\mu^2)$, is evolved from a scale $\mu_0^2$ up to $\mu^2$ using LO DGLAP evolution without quarks. The initial gluon density at the scale $\mu_0^2$ is taken in the form $ xg(x,\mu_0^2) = A_g\,x^{-\lambda_g}\,(1-x)^{5.6}$. The proton shape function $T(b)$ is normalized so that $\int\!\dif^2\vec{b}\;T(b) = 1$ and one considers a Gaussian form for $T(b)$, that is, $ T_G(b) = \frac{1}{2\pi B_G}\mathrm{e}^{-\frac{b^2}{2B_G}}$ where $B_G=4$ GeV$^{-2}$ \cite{KMW}.

At high energies, a power fit can be also performed for the WATT and b-SAT models. For WATT impact parameter model, we obtain the parameters $A= 4.3$ pb and $\alpha = 0.37$. In the b-SAT model we have the values $A= 2$ pb and $\alpha = 0.57$. We clearly verify a distinct energy dependence, which depends on the characteristics features of the phenomenological models.  The main point is the value of the parameter $\lambda$ entering at the saturation scale. The WATT model gives the softer energy dependence, which comes from the small value of $\lambda = 0.12$ in the saturation scale, $Q_{\mathrm{sat}}^2(x,t=0)\propto x^{\lambda}$. On  the other hand, in the MPS model one has $\lambda = 0.22$ (a factor two larger than in WATT model) and the QCD evolution makes the effective $\lambda$ value for b-SAT model to be large. The different overall normalization at energies $W\leq 100$ GeV for the MPS model is probably result of distinct behavior towards low energies.

Let us discuss the connection of the present calculation with the conventional partonic description of timelike Compton scattering. Presently, the photoproduction of a heavy timelike photon which decays into a lepton pair, $\gamma p \rightarrow \ell^+\ell^-\,p$, is computed to leading twist and at Born level \cite{Diehl}. In the Bjorken limit, i.e. large $Q^2$, the QCD factorization theorem represents the amplitude  by the convolution of hard scattering coefficients, calculable in perturbation theory, and generalized parton distributions (GPDs), which describe the nonperturbative physics of the process. GPDs are universal, process-independent, functions that contain information on parton distributions and correlations in hadrons and in matrix elements describing transitions between different hadrons.  For instance, in Ref. \cite{Diehl}, the quark handbag diagrams (leading order in $\alpha_s$) and simple models of the relevant GPDs are used to estimate the cross section and the angular asymmetries for lepton pair production. We remark that such a calculation can not be compared with ours since an explicit analysis using ${\cal{O}}(\alpha_s)$ accuracy, one-loop corrections to the quark handbag diagrams and other diagrams involving gluon distributions is currently unknown. Finally, it should be noticed that the Bethe-Heitler (BH) mechanism contributes at the amplitude level to the physical process of photoproduction of heavy lepton pair. It is known that the BH contribution (and its interference with TCS) is large in contrast to timelike Compton scattering and a careful numerical analysis is deserved.

Finally, we present  a study on the TCS process for scattering off nucleus. Namely, we focus on the process $\gamma A \rightarrow \gamma^*(\rightarrow \ell^+\ell^-)\,A$. In this case, we are considering the coherent contribution to nuclear TCS, where the nucleus remains intact. The implementation of nuclear effects in such a process is relatively   simple within the color dipole formalism in the small-$x$ region. The usual procedure is to consider the Glauber-Gribov formalism for nuclear absorption. In phenomenological models in which geometric scaling is present (as in MPS and WATT) the extrapolation for a nucleus target is simplified. Therefore, here we will rely on the geometric scaling property \cite{gsincl,travwaves} of the saturation models: such a scaling  means that  the total $\gamma^* p$
cross section at large energies is not a function of the two
independent variables $x$ and $Q$, but is rather a function of the
single variable $\tau_p = Q^2/Q_{\mathrm{sat}}^2(x)$ as shown
in Ref. \cite{gsincl}. That is, $\sigma_{\gamma^*p}(x,Q^2)=\sigma_{\gamma^*p}(\tau_p)$. In Refs. \cite{travwaves} it was shown that the geometric scaling observed in experimental data can be understood theoretically in the context of non-linear QCD evolution with fixed and running coupling. Recently, the high energy $l^{\pm}p$,  $pA$ and $AA$ collisions have been related through geometric scaling \cite{Armesto_scal}. Within the color dipole picture and making use of a rescaling of the impact parameter of the $\gamma^*h$ cross section in terms of hadronic target radius $R_h$, the nuclear dependence of the $\gamma^*A$ cross section is absorbed in the $A$-dependence of the saturation scale via geometric scaling. The relation reads as $\sigma^{\gamma^*A}_{tot}(x,Q^2)  =  \kappa_A\,\sigma^{\gamma^*p}_{tot}\,(Q_{\mathrm{sat},p} \rightarrow Q_{\mathrm{sat},A})$, where $\kappa_A = (R_A/R_p)^2$. The nuclear saturation scale was assumed to rise with the quotient of the transverse parton densities to the power $\Delta \approx 1$ and $R_A$ is the nuclear radius, $Q_{\mathrm{sat},A}^2=(A/\kappa_A)^{\Delta}\,Q_{\mathrm{sat},p}^2$. This assumption successfully describes small-$x$ data for $ep$ and $eA$ scattering using $\Delta =1.26$ and a same scaling curve for the proton and nucleus \cite{Armesto_scal}.

\begin{figure}[t]
\includegraphics[scale=0.47]{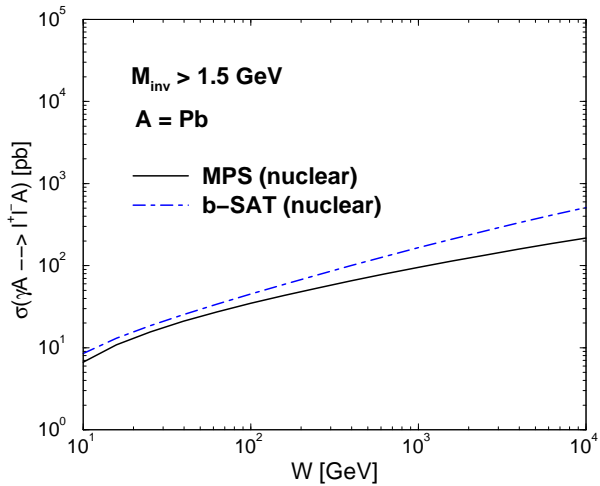}
\caption{(Color online)  The photonuclear cross section per nucleon as a function of energy for a lead nucleus (see text).}
\label{fig:4}
\end{figure}

Following the arguments above, we will replace $R_p \rightarrow R_A$ in Eq. (\ref{sigdipt}) and also $Q_{\mathrm{sat},p}^2 (x,t=0)\rightarrow (AR_p^2/R_A^2)^{\Delta}\,Q_{\mathrm{sat},p}^2(x,t=0)$ in Eq. (\ref{qsatt}). In case of $\Delta=1$ the previous replacement becomes the usual assumption for the nuclear saturation scale, $Q_{\mathrm{sat},A}^2=A^{1/3}\,Q_{\mathrm{sat},p}^2$. For simplicity, we replace the form factor $F(t)=\exp (-B|t|)$ in Eq. (\ref{sigdipt}) by the corresponding nuclear form factor $F_A(t)= \exp (-\frac{R_A^2}{6}|t|)$. In a careful calculation, a realistic nuclear form factor (for instance, hard sphere profile)  is needed. The corresponding phenomenological results are presented in Fig. \ref{fig:4}, where we have the integrated cross section ($M_{\ell^+\ell^-}\geq 1.5$ GeV and $|t|\leq 1$ GeV$^2$) per nucleon as a function of energy. We consider $A=208$ (Lead), which is relevant for electromagnetic interactions at $AA$ collisions at the LHC. The nuclear version of MPS model (using geometric scaling property) is represented by the solid line. At high energies ($W\geq 100$ GeV) the numerical result can be parameterized as $\sigma_{\mathrm{MPS}}(\gamma A\rightarrow \ell^+\ell^-\,A)=6.1\mathrm{pb}\,(W/W_0)^{0.39}$. For the b-SAT model, we just replace the proton shape function $T(b)$ in Eq (\ref{bsatelem}) by the corresponding nuclear profile $T_A(b)$ (Wood-Saxon). The result is presented by dot-dashed line and at high energies is parametrized as $\sigma_{\mathrm{bSAT}}(\gamma A\rightarrow \ell^+\ell^-\,A)=5\mathrm{pb}\,(W/W_0)^{0.51}$. Accordingly, we see a larger suppression in the MPS model than in b-SAT, which is directly related to the nuclear saturation scale at each model. Absorption is evident in MPS model, where the effective power on energy has diminished in the nuclear case.

Here, some comments are in order. The approximation we have considered to compute the photonuclear cross section should be taken with due care. In the small-$x$ limit, within the GPDs formalism the ratio of the real parts of the nuclear to nucleon amplitudes  exclusive processes (like DVCS) has a very unexpected behavior as a function of $x$, which is distinct from the corresponding ratio of imaginary parts. This is the reason we have disregarded the real part contribution in our numerical calculations. Another important aspect is the interplay between the coherent and incoherent contributions to the nuclear TCS. When the recoiled nucleus is not detected, the measurements of TCS observables with nuclear targets involves both contributions and their interference. In the coherent scattering the nuclear target stays intact and it dominates at small-$t$. On the other hand, in the incoherent process the initial nucleus $A$ transforms into the system of $A-1$ spectator nucleon and one interacting nucleon (it dominates at large-$t$). We have considered in present study the situation when the recoiled nucleus is detected and the $\gamma A\rightarrow \gamma^* A$ cross section is purely coherent. For a scenario with excited or break up nucleus, the cut $|t|\ll 1$ GeV$^2$  could prevent us of considering the incoherent contribution. The coherent contribution dominates the nuclear cross section at small-$t$.

The present calculations are input for the exclusive photoproduction of dileptons in electromagnetic interactions in $pp$ and nucleus-nucleus collisions. Such processes are characterized by the photon - hadron interaction, with the photon stemming from the electromagnetic field
of one of the two colliding hadrons (For recent reviews see Ref. \cite{YRUPC}).  The total cross section for the $h\,h\rightarrow h\otimes \ell^+\ell^- \otimes h$ process is given by
\begin{eqnarray}
\frac{d^2\sigma}{d\omega dM_{\ell^+\ell^-}^2} (h h \rightarrow  \ell^+\ell^- \,h h)\, = 2 \frac{dN_{\gamma}}{d\omega}\frac{d\sigma }{dM_{\ell^+\ell^-}^2} \left(\gamma h \rightarrow \ell^+\ell^- \,h\right) ,\nonumber
\label{sigAA}
\end{eqnarray}
where
$\omega$ is the photon energy   and $ \frac{dN_{\gamma}(\omega)}{d\omega}$ is the equivalent flux of photons from a charged hadron. Moreover, $\gamma_L$ is the Lorentz boost  of a single beam,  $W_{\gamma h}^2=2\,\omega\sqrt{S_{\mathrm{NN}}}$  and
$\sqrt{S_{\mathrm{NN}}}$ is  the c.m.s energy of the
hadron-hadron system \cite{YRUPC}. This process is characterized by small momentum transfer and energy loss, which implies that the outgoing hadrons should be detected in the forward regions of detectors. A rough estimation for $pp$ collisions  gives dozens of pb at Tevatron and a hundred of pb at the LHC (integrated over invariant mass $M_{\mathrm{inv}}>1.5$ GeV). In the $AA$ the photon flux is enhanced by a factor $\propto Z^2$, then we roughly expect the cross sections reach dozens of nb at $PbPb$ collisions at the LHC. The exclusive photoproduction of dileptons from TCS has similar final state configuration than the QED process $\gamma \gamma \rightarrow \ell^+\ell^-$. However, detailed estimations deserve further studies as the correct time-like kinematics and realistic nuclear profiles, which we postpone for a future publication.

\section{Summary}

Using the color dipole formalism, we studied the timelike Compton scattering. Such an approach is robust in describing a wide class of exclusive processes measured at DESY-HERA and at the experiment CLAS (Jeferson Lab.), like meson production, diffractive DIS and DVCS. Our investigation is complementary to conventional partonic description of TCS, which considers quark handbag diagrams (leading order in $\alpha_s$) and simple models of the relevant GPDs. In particular, the results could be compared to pQCD diagrams involving gluon distributions which are currently unknown. Using current phenomenology for the elementary dipole-hadron scattering, we estimate the order of magnitude of the exclusive photoproduction of lepton pairs. In the present analysis we also investigate the photonuclear cross section, focusing  on the nuclear coherent scattering. The main results is that the cross section per nucleon reaches dozens of picobarns. At DESY-HERA (with $W=82$ GeV) the pure TCS contribution gives a cross section of 20 pb with an uncertainty of a few percents. It should be noticed that we consider space-like kinematics for numerical calculations and that the correct time-like kinematics will introduce corrections to our preliminary estimations.  These calculations are input in electromagnetic interactions in $pp$ and $AA$ collisions to measured at the LHC. We found that the exclusive photoproduction of lepton pairs in such reactions should be sizable.

\begin{acknowledgments}
 The author thanks the CERN Theory Division for hospitality and the living atmosphere of the Workshop on {\it High energy photon collisions at the LHC}, CERN 22-25 April 2008, where this work has started.  The clarifying comments of B. Pire are acknowledged.  The author is grateful to L. Motyka and G. Watt for useful informations. This work was supported by CNPq, Brazil.
\end{acknowledgments}

\end{document}